\documentclass[11pt]{article}
\usepackage{Blois,epsfig}

\def\be{\begin{equation}}
\def\ee{\end{equation}}
\def\bea{\begin{eqnarray}}
\def\eea{\end{eqnarray}}

\begin{document}
\vspace*{2cm}
\begin{center}
\Large{\textbf{XIth International Conference on\\ Elastic and Diffractive Scattering\\ Ch\^{a}teau de Blois, France, May 15 - 20, 2005}}
\end{center}

\vspace*{2cm}
\title{HIGH MASS DIFFRACTION AT THE LHC}

\author{C. ROYON}

\address{Service de physique des particules, CEA/Saclay,
  91191 Gif-sur-Yvette cedex, France \\ Fermi National Accelerator Laboratory, Batavia, USA}

\maketitle\abstracts{
We use a Monte Carlo implementation of recently developped 
models of exclusive diffractive $W$, top, Higgs and stop productions
to assess the sensitivity of the LHC experiments.}

\section{Theoretical framework}

Let us introduce the model \cite{bialas} we shall use for describing 
exclusive SUSY Higgs bosons and stop pair production in double diffractive 
production. In \cite{bialas}, the diffractive mechanism is based on two-gluon 
exchange between the 
two incoming protons. The soft pomeron is seen as a  pair of gluons 
non-perturbativelycoupled  to the proton. One of the gluons is then coupled 
perturbatively to the hard process, either the SUSY Higgs bosons, or 
the $\tilde t \bar{\tilde t}$ pair, while the other one plays the r\^ole of a 
soft screening of colour, allowing for diffraction to occur.
The corresponding cross-sections for Higgs bosons and $\tilde t \bar{\tilde t}$ 
production read:

\begin{eqnarray}
 d\sigma_{h}^{exc}(s) &=& C_{h}\left(\frac{s}{M_{h}^{2}}\right)^{2\epsilon} 
\delta\left(\xi_{1}\xi_{2}-\frac{M_{h}^{2}}{s}\right)
\prod_{i=1,2} \left\{ d^{2}v_{i} \frac{d\xi_{i}}{1-\xi_{i}} \right.
\left. \xi_{i}^{2\alpha'v_{i}^{2}} \exp(-2\lambda_{h} v_{i}^{2})\right\} 
\sigma (g g \rightarrow h)
 \nonumber \\
d\sigma_{\tilde t \tilde{\bar{t}}}^{exc}(s) &=& C_{\tilde t \tilde {\bar{t}}} 
\left(\frac{s}{M_{\tilde t \tilde{\bar{t}}^{2}}}\right)^{2\epsilon}
\delta\left( \sum_{i=1,2} (v_{i} + k_{i}) \right)
\prod_{i=1,2} \left\{ d^{2}v_{i} d^{2}k_{i} d\xi_{i} \right.
d\eta_{i}\  \xi_{i}^{2\alpha'v_{i}^{2}}\! 
\left. \exp(-2\lambda_{\tilde t\tilde{\bar{t}}} v_{i}^{2})\right\} {\sigma} 
(gg\to {\tilde t \tilde{\bar{t}}}\ ) \nonumber
\label{exclusif}
\end{eqnarray}
where, in both equations,  the variables $v_{i}$ and $\xi_{i}$ respectively 
denote the transverse
momenta and fractional momentum losses of the outgoing protons. In the second 
equation, 
$k_{i}$ and $\eta_{i}$ are respectively the squark  transverse 
momenta and  rapidities. $\sigma (g g \rightarrow H),  {\sigma} (gg\to {\tilde t 
\tilde{\bar{t}}}\ )$ are the hard  production cross-sections which are given 
later on. The model normalisation constants  $C_{h}, C_{\tilde t \tilde 
{\bar{t}}}$ are fixed from the fit to dijet diffractive production.

In the model,  the soft pomeron 
trajectory is
taken 
from the standard 
Donnachie-Landshoff   parametrisation \cite{donnachie},
 namely $\alpha(t) = 1 + 
\epsilon + \alpha't$, with
$\epsilon \approx 0.08$ and $\alpha' \approx 0.25 
\mathrm{GeV^{-2}}$. 
$\lambda_{h}, \lambda_{\tilde t \tilde{\bar{t}}}$ are  
kept as in  the original paper \cite{bialas} for the SM Higgs and $q 
\bar q$ pairs.  Note that, in this model, the 
strong (non perturbative) 
coupling constant is fixed to a reference value 
$G^2/4\pi,$ which will be taken 
from the fit to the observed centrally produced diffractive dijets.

In order to select exclusive diffractive states,  it is required to take into 
account the corrections from soft hadronic scattering. Indeed, the soft 
scattering  between incident particles tends to mask the genuine
hard diffractive interactions at 
hadronic colliders. The formulation of this 
correction \cite{alexander} to the
scattering amplitudes consists in considering a gap 
survival  probability.
The correction factor is commonly evaluated to be of order $0.03$ for the QCD 
exclusive diffractive processes at the LHC.

More details about the theoretical model and its phenomenological
applications can be found in Refs. \cite{ourpap} and \cite{us}. In the following,
we use the BL model for exclusive Higgs production recently implemented in
a Monte-Carlo generator \cite{ourpap}. 

\section{Experimental context}

The analysis is based on a fast simulation of the CMS detector at the LHC
(similar results would be obtained using the ATLAS simulation).
The calorimetric coverage of the CMS experiment ranges up to a pseudorapidity 
of $|\eta|\sim 5$. 
The region devoted
to precision measurements lies within $|\eta|\leq 3$, with a typical 
resolution on jet energy measurement of $\sim\!50\%
/\sqrt{E}$,
where $E$ is in GeV, and a granularity in pseudorapidity and azimuth of 
$\Delta\eta\times\Delta\Phi \sim 0.1\times 0.1$. 

In addition to the central CMS detector, the existence of roman pot detectors
allowing to tag diffractively produced protons,
located on both $p$ sides, is assumed \cite{helsinki}. The $\xi$ acceptance and 
resolution have been derived for each device using a complete simulation
of the LHC beam parameters. The combined $\xi$ acceptance is $\sim 100\%
$ 
for $\xi$ ranging from $0.002$ to $0.1$, where
$\xi$ is 
the proton fractional momentum loss. The acceptance limit of the device 
closest to the interaction point
is $\xi > \xi_{min}=$0.02. 

In exclusive double Pomeron exchange, the mass of the central 
heavy object is given by $M^2 = \xi_1\xi_2 s$, where $\xi_1$ and $\xi_2$ are
the proton fractional momentum losses measured in the roman pot detectors.
\section{Existence of exclusive events}
The question arises if exclusive events exist or not since they have never been
observed so far. The D\O\ and CDF experiments 
at the Tevatron (and the LHC experiments) are ideal places to look for
exclusive events in dijet or $\chi_C$ channels for instance
where exclusive events are expected to occur at high dijet mass
fraction.
So far, no evidence of the existence of exclusive events has been found.
A nice way to show the existence of such events would be to study the
correlation between the gap size measured in both $p$ and $\bar{p}$ directions
and the value of $log 1/\xi$ measured using roman pot detectors, which can be
performed in the D\O\ experiment. The gap size between the
pomeron remnant and the protons detected in roman pot detector 
is of the order of 
$log 1/\xi$ for usual diffractive events (the measurement giving a slightly
smaller value to be in the acceptance of the forward detectors) while
exclusive events show a much higher value for the rapidity gap since the gap
occurs between the jets (or the $\chi_C$) and the proton detected in roman
pot detectors (in other words, there is no pomeron remnant)
\footnote{To distinguish between pure exclusive and
quasi-exclusive events, other observables such as
the ratio of the cross sections of double diffractive
production of diphoton and dilepton, or the $b$-jets to all jets 
are needed \cite{us}.}. Another observable leading
to the same conclusion would be the correlation between $\xi$ computed
using roman pot detectors and using only the central detector.

\section{Results on diffractive Higgs production}
Results are given in Fig. 1 for a Higgs mass of 120 GeV, 
in terms of the signal to background 
ratio S/B, as a function of the Higgs boson mass resolution.

In order to obtain an S/B of 3 (resp. 1, 0.5), a mass resolution of about
0.3 GeV (resp. 1.2, 2.3 GeV) is needed. The forward detector design of 
\cite{helsinki} 
claims a resolution of about 2.-2.5 GeV, which leads to a S/B of about 
0.4-0.6. Improvements in this design
would increase the S/B ratio as indicated on the figure.
As usual, this number is enhanced by a large factor if one considers 
supersymmetric Higgs boson 
production with favorable Higgs or squark field mixing parameters.

The cross sections obtained after applying the survival probability of 0.03 at
the LHC as well as the S/B ratios are given in Table \ref{sb} if one assumes a
resolution on the missing mass of about 1 GeV (which is the most optimistic
scenario). The acceptances of the roman pot detectors as well as the simulation
of the CMS detectors have been taken into account in these results. 

Let us also notice that the missing mass method will allow to perform a $W$ 
mass measurement using exclusive (or quasi-exclusive) $WW$ 
events in double Pomeron exchanges, and QED processes. The advantage of the
QED processes is that their cross section is perfectly known and that this
measurement only depends on the mass resolution and the roman pot acceptance.
In the same way, it is possible to measure the mass of the top quark in
$t \bar{t}$ events in double Pomeron exchanges.

\begin{table}
\begin{center}
\begin{tabular}{|c||c|c|c|c|c|} \hline
$M_{Higgs}$& cross & signal & backg. & S/B & $\sigma$  \\
 & section &  & & & \\
\hline\hline
120 & 3.9 & 27.1 & 28.5 & 0.95 & 5.1  \\
130 & 3.1 & 20.6 & 18.8 & 1.10 & 4.8  \\
140 & 2.0 & 12.6 & 11.7 & 1.08 & 3.7  \\ 
\hline
\end{tabular}
\caption{Exclusive Higgs production cross section for different Higgs masses,
number of signal and background events for 100 fb$^{-1}$, ratio, and number of
standard deviations ($\sigma$).}
\label{sb}
\end{center}
\end{table}

The diffractive SUSY Higgs boson production cross section is noticeably enhanced 
at high values of $\tan \beta$ and since we look for Higgs decaying into $b
\bar{b}$, it is possible to benefit directly from the enhancement of the cross
section contrary to the non diffractive case. A signal-over-background up to a
factor 50 can be reached for 100 fb$^{-1}$ for $\tan \beta \sim 50$
\cite{lavignac}. We give in Figure 2 the
signal-over-background ratio for different values of $\tan \beta$ for a Higgs
boson mass of 120 GeV.

%\begin{figure}[!thb]
%\vspace*{7.0cm}
%\begin{center}
%\special{psfile=signaloverback.ps voffset=-50 vscale=45
%hscale= 50 hoffset=-35 angle=0}
%%\centerline{\epsfxsize=5.9in\epsfbox{signaloverback.ps}}
%\vspace{1.cm}
%\caption[*]{Standard Model Higgs boson signal to background ratio as a function of the resolution on 
%	the missing mass, in GeV. This figure assumes a Higgs
%	boson mass of 120 GeV. }
%\end{center}
%\end{figure}

\begin{figure}[htb]
\begin{center}
%\begin{minipage}[t]{70mm}
\includegraphics[width=11.5cm,clip=true]{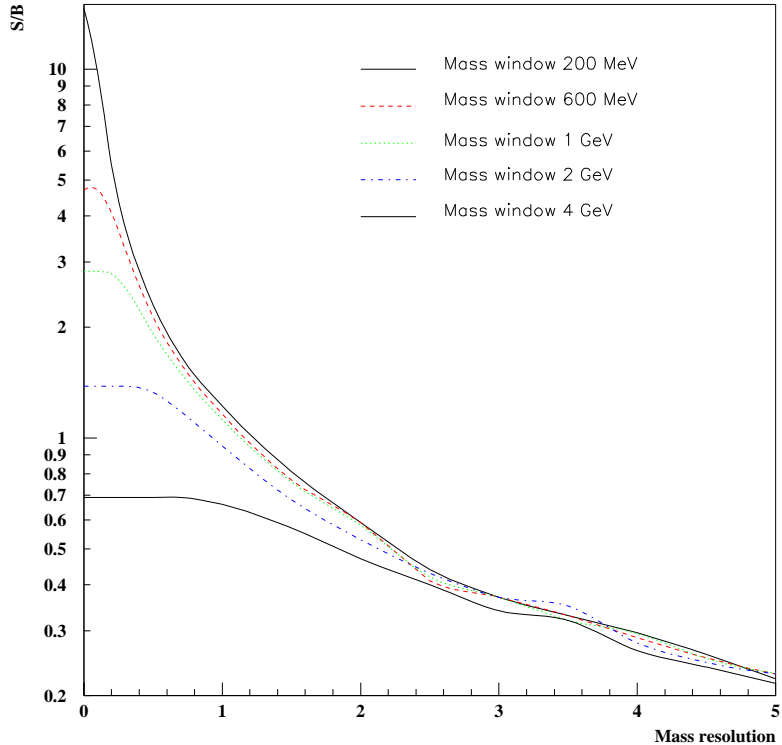}
%\end{minipage}
\caption{Standard Model Higgs boson signal to background ratio as a function 
  of the resolution on
        the missing mass, in GeV. This figure assumes a Higgs
        boson mass of 120 GeV.}
\end{center}
\end{figure}

\begin{figure}[htb]
\begin{center}
%\begin{minipage}[t]{70mm}
\includegraphics[width=11.5cm,clip=true]{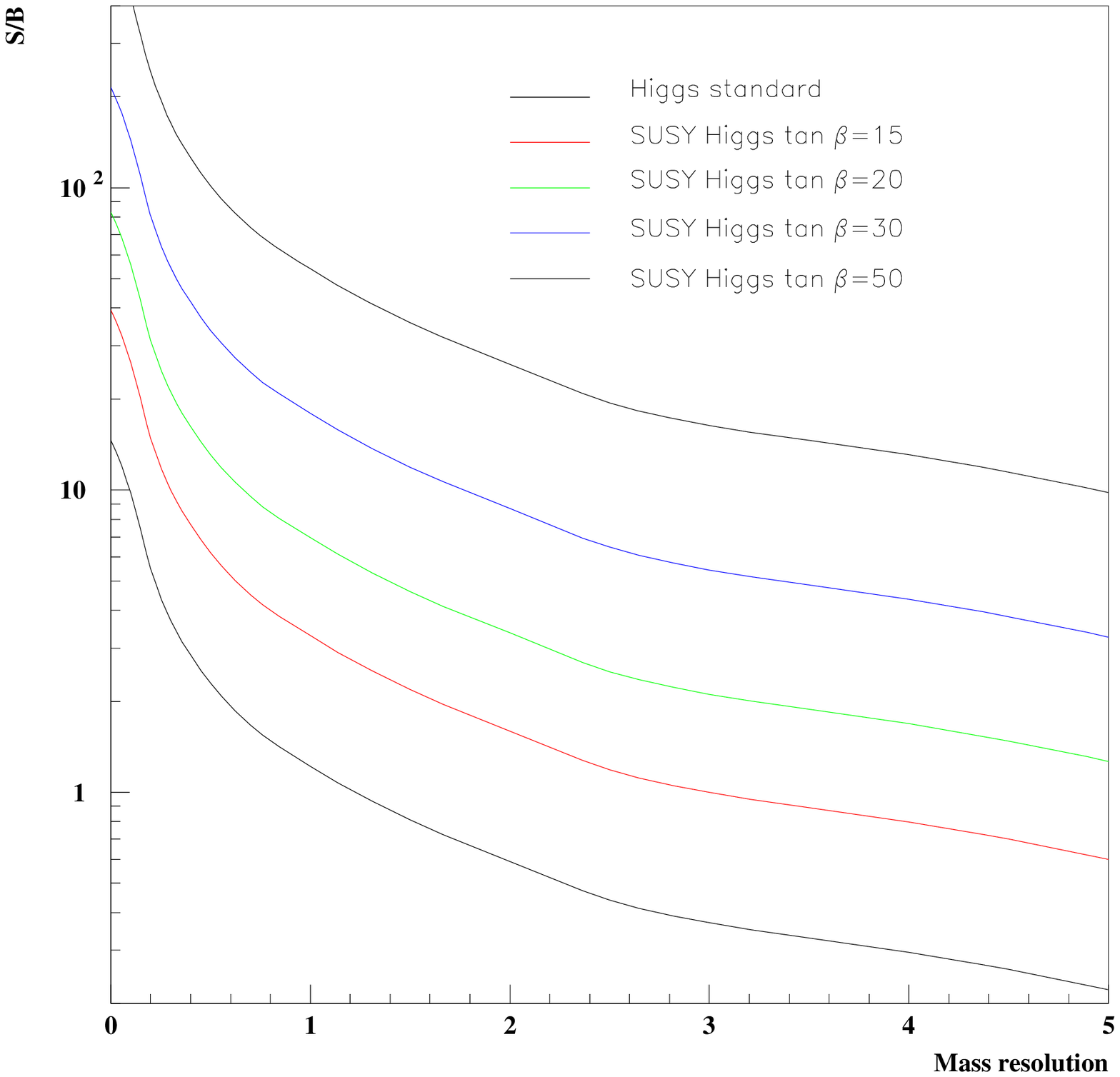}
%\end{minipage}
\caption{SUSY Higgs boson signal to background ratio as a function 
  of the resolution on
        the missing mass, in GeV. This figure assumes a Higgs
        boson mass of 120 GeV.}
\end{center}
\end{figure}

\section{Threshold scan method: $W$, top and stop mass measurements}
We propose a new method to measure heavy particle properties via double 
photon and double pomeron exchange (DPE), at the LHC \cite{ushiggs}. In this category of events, the heavy objects 
are produced in pairs, whereas the beam particles
often leave the interaction region intact, and can be measured using very forward detectors.

Pair production of $WW$ bosons and top quarks in QED and  double pomeron exchange are described in detail in this section. 
$WW$ pairs are produced in photon-mediated processes, which are exactly calculable in QED. There is 
basically no uncertainty concerning the possibility of measuring these processes
at the LHC. On the contrary, $t \bar{t}$ events, produced in 
exclusive double pomeron exchange, suffer from theoretical uncertainties since 
exclusive diffractive production is still to be observed at the Tevatron, 
and other models lead to different cross sections, and thus to a different
potential for the top quark mass measurement. However, since the exclusive 
kinematics are simple, the model dependence will be essentially reflected by 
a factor in the effective luminosity for such events.

\subsection{Explanation of the methods}
We study two different methods to reconstruct the mass of heavy objects
double diffractively produced at the LHC. The method is
based on a fit to the turn-on point of the missing mass distribution at 
threshold. 

One proposed method (the ``histogram'' method) corresponds to the comparison of 
the mass distribution in data with some reference distributions following
 a Monte Carlo simulation of the detector with different input masses
corresponding to the data luminosity. As an example, we can produce 
a data sample for 100 fb$^{-1}$ with a top mass of 174 GeV, and a few 
MC samples corresponding to top masses between 150 and 200 GeV by steps of. 
For each Monte Carlo sample, a $\chi^2$ value corresponding to the 
population difference in each bin between data and MC is computed. The mass point 
where
the $\chi^2$ is minimum corresponds to the mass of the produced object in data.
This method has the advantage of being easy but requires a good
simulation of the detector.

The other proposed method (the ``turn-on fit'' method) is less sensitive to the MC 
simulation of the
detectors. As mentioned earlier, the threshold scan is directly sensitive to
the mass of the diffractively produced object (in the $WW$W case for instance, it
is sensitive to twice the $WW$ mass). The idea is thus to fit the turn-on
point of the missing mass distribution which leads directly to the mass 
of the produced object, the $WW$ boson. Due to its robustness,
this method is considered as the ``default" one in the following.

\subsection{Results}

To illustrate the principle of these methods and their achievements,
we  apply them to the 
$WW$ boson and the top quark mass measurements in the
following, and obtain the reaches at the LHC. They can be applied to other 
threshold scans as well.
The precision of the $WW$ mass measurement (0.3 GeV for 300 fb$^{-1}$) is not competitive with other 
methods, but provides a very precise calibration 
of the roman pot detectors. The precision of
the top mass measurement is however competitive, with an expected precision 
better than 1 GeV at high luminosity. The resolution on the top mass is given
in Fig. 3 as a function of luminosity for different resolutions of the roman
pot detectors.

The other application is to use the so-called ``threshold-scan method"
to measure the stop mass in {\it exclusive} events. The idea is straightforward: 
one
measures the turn-on point in the missing mass distribution at about twice
the stop mass. After taking into account the stop width, we obtain a resolution
on the stop mass of 0.4, 0.7 and 4.3 GeV for a stop mass of 174.3, 210 and 393
GeV for a luminosity (divided by the signal efficiency) of 100 fb$^{-1}$. We
notice that one can expect to reach typical mass resolutions which can be obtained at a linear
collider. The process is thus similar to  those at linear colliders (all final 
states
are detected) without the initial state radiation problem. 

The caveat is  of course that production via diffractive 
{\it exclusive} processes is model dependent, and definitely needs
the Tevatron data to test the models. It will allow to determine more precisely 
the production cross section by testing and measuring at the Tevatron the jet 
and photon production for high masses and high dijet or diphoton mass fraction.

\begin{figure}[htb]
%\begin{minipage}[t]{70mm}
\includegraphics[width=12.5cm,clip=true]{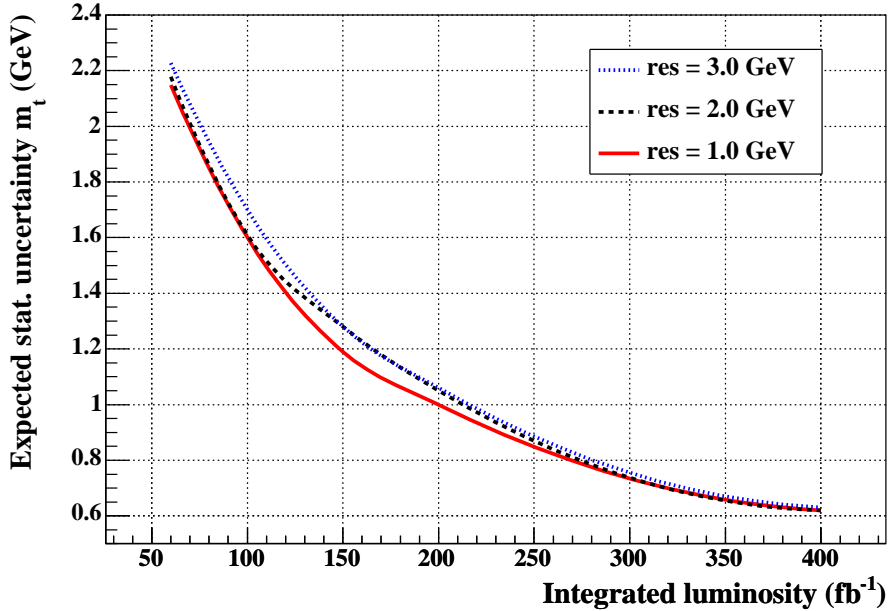}
%\end{minipage}
\caption{Expected statistical precision of the top mass
    as a function of the integrated luminosity for various resolutions
    of the roman pot detectors (full line: resolution of 1 GeV, dashed line: 2
    GeV, dotted line: 3 GeV).}
\end{figure}

\section*{Acknowledgments}
There results come from a fruitful collaboration with M. Boonekamp, J. Cammin,
S. Lavignac and R. Peschanski. The author also thanks the organisers for
financial support.


\section*{References}
\begin{thebibliography}{99}

\bibitem{bialas} 
A. Bialas, P.V. Landshoff, Phys. Lett.  {\bf B256} (1990) 540.


\bibitem{donnachie}
A. Donnachie, P.V. Landshoff, Phys. Lett. {\bf B296} (1992) 227.


\bibitem{alexander} A. Kupco, R. Peschanski, C.Royon,
Phys. Lett. {\bf B606} (2005) 139.

\bibitem{ourpap}  M. Boonekamp, R. Peschanski, and C. Royon,
Phys. Lett. {\bf B598} (2004) 243.



\bibitem{us} 
M. Boonekamp, R. Peschanski, C. Royon, Phys. Rev. Lett.  {\bf  87 } 
(2001) 251806;
M. Boonekamp, A. De Roeck, R. Peschanski, C. Royon, Phys. Lett.  {\bf  B550} 
(2002) 93;
M. Boonekamp, R. Peschanski, C. Royon, Nucl. Phys. {\bf B669} (2003) 277, 
Err-ibid {\bf B676} (2004) 493;
for a general review see C. Royon, Mod. Phys. Lett. {\bf A18} (2003) 2169.




\bibitem{helsinki} 
J. Kalliopuska, T. M\"aki, N. Marola, R. Orava, K. \"Osterberg, 
M. Ottela, HIP-2003-11/EXP.




\bibitem{ushiggs} M. Boonekamp, J. Cammin, R. Peschanski, C. Royon,
hep-ph/0504199.

\bibitem{lavignac} M. Boonekamp, J. Cammin, S. Lavignac, R. Peschanski,
C. Royon, hep-ph/0506275.



\end{thebibliography}
\end{document}